\documentclass[12pt]{article}

\newcommand{\VersionInformation}{}  
\InputIfFileExists{Debug}{}{}

\usepackage[utf8]{inputenc}
\usepackage{amsmath}
\usepackage{amsthm}
\usepackage{amsfonts}          
\usepackage{amssymb}           
\usepackage[mathscr]{eucal}
\usepackage{graphicx}
\usepackage{color}
\usepackage[normalem]{ulem}
\usepackage{hypbmsec}
\usepackage{fancyvrb}
\usepackage{sepnum}
\usepackage{xspace}

\usepackage{rotating}
\usepackage[isu,bf]{caption}
\usepackage{multirow}

\newlength{\xtrawidth}
\newlength{\xtraheight}
\ifx\DEBUG\EMPTYMACRO
\else
\fi

\def\clap#1{\hbox to 0pt{\hss#1\hss}}

\makeatletter
  \def\adots{\mathinner{\mkern2mu\raise\p@\hbox{.}
      \mkern2mu\raise4\p@\hbox{.}\mkern1mu
      \raise7\p@\vbox{\kern7\p@\hbox{.}}\mkern1mu}}
\makeatother

\newcommand{\eqdef}{%
  \mathrel{\lower.1mm
    \hbox{$\stackrel{\lower.424ex\hbox{\scriptsize def}}{=}$}}
}

\newcommand{\R}{\ensuremath{{\mathbb{R}}}}
\newcommand{\C}{\ensuremath{{\mathbb{C}}}}

\newcommand{\Z}{\mathbb{Z}}
\newcommand{\CP}{{\ensuremath{\mathop{\null {\mathbb{P}}}\nolimits}}}

\newcommand{\tors}{\ensuremath{\text{tors}}}

\DeclareMathOperator{\diff}{d\!}

\DeclareMathOperator{\rank}{rank}
\DeclareMathOperator{\codim}{codim}

\DeclareMathOperator{\Hom}{Hom}

\newcommand{\Osheaf}{\ensuremath{\mathscr{O}}}

\newcommand{\dual}{\ensuremath{\vee}}

{\end{list}}

{\everymath{\displaystyle\everymath{}}\array}%
{\endarray}

\pdfoutput=1

\newcommand{\Bsmooth}{{\hat{B}}}
\newcommand{\Bmed}{B}
\newcommand{\Bsing}{{\uline{B}}}

\newcommand{\Dsmooth}{{\hat{D}}}
\newcommand{\Dmed}{D}
\newcommand{\Dsing}{{\uline{D}}}

\newcommand{\Ysmooth}{{\hat{Y}}}
\newcommand{\Ysing}{{\uline{Y}}}

\newcommand{\dsmooth}{{\hat{d}}}

\newcommand{\dsing}{{\uline{d}}}

\newcommand{\Rsmooth}{{\hat{R}}}
\newcommand{\rsmooth}{{\hat{r}}}

\newcommand{\zsmooth}{{\hat{z}}}
\newcommand{\zmed}{z}
\newcommand{\zsing}{{\uline{z}}}

\newcommand{\pismooth}{\ensuremath{\hat{\pi}}}
\newcommand{\pising}{\ensuremath{\uline{\pi}}}

\DeclareMathOperator{\conv}{conv}
\newcommand{\Kc}{\ensuremath{\mathcal{K}}}

\newcommand{\Sigmaone}{\ensuremath{\Sigma^{(1)}}}

\begin{document}
\begin{titlepage}
  \vspace*{-2cm}
  \VersionInformation
  \hfill
  \parbox[c]{5cm}{
    \begin{flushright}
      DIAS-STP 10-11
    \end{flushright}
  }
  \vspace*{2cm}
  \begin{center}
    \Huge 
    Discrete Wilson Lines in F-Theory
  \end{center}
  \vspace*{8mm}
  \begin{center}
    \begin{minipage}{\textwidth}
      \begin{center}
        \sc 
        Volker Braun
      \end{center}
      \begin{center}
        \textit{
          Dublin Institute for Advanced Studies\hphantom{${}^1$}\\
          10 Burlington Road\\
          Dublin 4, Ireland
        }
      \end{center}
      \begin{center}
        \texttt{Email: vbraun@stp.dias.ie}
      \end{center}
    \end{minipage}
  \end{center}
  \vspace*{\stretch1}
  \begin{abstract}
    F-theory models are constructed where the $7$-brane has a
    non-trivial fundamental group. The base manifolds used are a toric
    Fano variety and a smooth toric threefold coming from a reflexive
    polyhedron. The discriminant locus of the elliptically fibered
    Calabi-Yau fourfold can be chosen such that one irreducible
    component it is not simply connected (namely, an Enriques surface)
    and supports a non-Abelian gauge theory.
  \end{abstract}
  \vspace*{\stretch1}
\end{titlepage}
\tableofcontents
\listoffigures 	
\listoftables 	

\section{Introduction}
\label{sec:into}

F-theory~\cite{Vafa:1996xn} is a way to use geometry as a tool to
understand certain compactifications of string theory that are
otherwise not entirely geometric~\cite{Sen:1996vd}. It uses an
auxiliary elliptically fibered Calabi-Yau fourfold, not to be confused
with the space-time manifold to study string theory in a regime away
from any known weakly-coupled perturbative
description. Recently~\cite{Beasley:2008dc, Beasley:2008kw} a
particular model building Ansatz has been suggested where the GUT
gauge group arises from a $7$-brane wrapped on a contractible del
Pezzo surface. Various models~\cite{Donagi:2008ca, Marsano:2009ym,
  Blumenhagen:2009yv, Chen:2010ts, Chen:2010tp, Blumenhagen:2010at}
and more have been constructed along these lines.

One key feature of this Ansatz is that the scales of gravity and gauge
physics can be decoupled as one can decompactify the Calabi-Yau
manifold without changing the del Pezzo surface. However, the price
one has to pay for this is that the usual way of GUT symmetry breaking
in string theory, namely the Hosotani mechanism\cite{Hosotani1,
  Hosotani2} using discrete Wilson lines, no longer works: All del
Pezzo surfaces are simply connected. Alternatives have been
developed~\cite{Beasley:2008dc, Beasley:2008kw, Blumenhagen:2008zz},
but require one to give a vacuum expectation value to fields locally
and not just make global non-trivial identifications. Turning on
fields locally then affects the running of the coupling constants and,
potentially, defocus the gauge coupling
unification~\cite{Donagi:2008kj, Blumenhagen:2008aw}.

In this paper I will advocate for a different Ansatz for GUT model
building and symmetry breaking in F-theory, namely, by wrapping the
GUT $7$-brane on a non-simply connected divisor in the base of the
elliptic fibration. This allows one to choose a globally non-trivial
identification of the gauge bundle while keeping it locally trivial,
breaking the GUT gauge group by the usual Hosotani mechanism. For what
its worth, this setup also implies that there is no gravity/gauge
theory decoupling limit.

Of course this raises the question of whether there are any such
divisors in threefolds that are suitable as bases for elliptically
fibered Calabi-Yau manifolds. In this paper I will answer this
question and work out a rather simple example of an Enriques surface
embedded into a toric threefold associated to a reflexive
3-dimensional polytope. There is nothing particularly unique about
this example; It just combines the most simple surface with $\Z_2$
fundamental group and the class of threefolds we are most used to work
with. All toric geometry computations used in this paper were done
using~\cite{ToricVarieties, Sage, GPS05}.

\section{Base Threefold}
\label{sec:base}

\subsection{Foreword}
\label{sec:fore}

An Enriques surface is a free quotient of a $K3$ surface by a
freely-acting holomorphic involution and is probably the best-known
example of a complex surface $S$ with fundamental group
$\pi_1(S)=\Z_2$. Its first Chern class $c_1(S)$ is the torsion element
in $H^2(Z,\Z)\simeq \Z^{10}\oplus \Z_2$, so it admits a Ricci-flat
metric but has no covariantly constant spinor.\footnote{Equivalently,
  no covariantly constant $(2,0)$-form.} Some, but not all, $K3$
surfaces can be realized~\cite{Nahm:1999ps} as quartics in
$\CP^3$. Somewhat unfortunately, the locus of quartic $K3$s and the
locus of $K3$ surfaces with an Enriques involution do not intersect in
the moduli space of smooth $K3$ surfaces. In other words, no smooth
quartic in $\CP^3$ carries an Enriques involution. Therefore, out of
necessity one is forced to look at singular (birational) models and
then resolve these singularities. This will be the central theme in
the following.

To explicitly construct and resolve the singularities, I will make
extensive use of toric geometry. However, before delving into these
technical details let me first give an overview. The basic idea is to
look at the following $\Z_4$ action on $\CP^3$, 
\begin{equation}
  g: \CP^3 \to \CP^3,\quad
  \big[x_0:x_1:x_2:x_3\big]
  \big[
  x_0:
  i x_1:
  i^2 x_2:
  i^3 x_3
  \big].
\end{equation}
The fixed point set of $g^2$ are the two disjoint rational curves
\begin{equation}
  \CP^1 \cup \CP^1 = 
  \big\{ x_0=x_2=0 \big\}
  \cup
  \big\{ x_1=x_2=0 \big\}
\end{equation}
and the fixed points of $g$ are the north and south poles on these
($4$ points altogether). A sufficiently generic $\Z_4$ invariant
quartic $q(x_0,x_1,x_2,x_3)$ is then a (singular) Enriques surface on
the quotient.\footnote{This construction is rather similar to the way
  to construct non-simply connected Calabi-Yau
  threefolds\cite{MR2282962, MR2122419, Braun:2010vc, Braun:2009qy,
    Candelas:2010ve}, except that I will not be looking at sections of
  the anti-canonical bundle (which would be Calabi-Yau).} The fastest
way to see this is to note that the would-be $(2,0)$-form
\begin{equation}
  \Omega^{(2,0)} 
  =
  \oint 
  \frac{ 
    \epsilon^{ijk\ell} x_i \diff x_j \wedge  \diff x_k \wedge \diff x_\ell 
  }{
    q(x_0,x_1,x_2,x_3)
  }
\end{equation}
is projected out by $g$.

Here is where this paper essentially begins, because so far we only
have a singular Enriques surface in an even more singular ambient
space. Clearly, one wants to resolve the singularities. The first step
is to resolve the curves of $\Z_2$ singularities, for which there is a
unique crepant resolution. Then one has to deal with the remaining $4$
$\Z_4$ singularities. By a happy coincidence, the above $\Z_4$
quotient of $\CP^3$ is itself a toric variety. Hence, the methods of
toric geometry can be applied and allow us to construct partial and
complete resolutions explicitly as toric varieties.

\subsection{Toric Geometry}
\label{sec:toric}

As a warm-up, I will first review some basic notions of toric
geometry. The defining data is a rational polyhedral fan in a lattice
$N\simeq \Z^d$, where $d$ is the complex dimension of the variety. A
fan $\Sigma$ is a finite set of cones $\sigma \in \Sigma$, closed
under taking faces. Often, the fan will be the cones over the faces of
a polytope. This is called the face fan of the polytope.

Amongst the different, but equivalent ways to define the corresponding
complex algebraic variety from the fan data, I will use the Cox
homogeneous coordinate~\cite{MR1299003} description in the
following. The basic idea is to associate one complex-valued
homogeneous coordinate to each ray (one-dimensional cone) of the
fan. Then one has to remove a codimension-$2$ or higher algebraic
subset and mod out generalized homogeneous rescalings. This
construction will be reviewed and applied in much more detail in
\autoref{sec:cox}. For now, let us just consider $\CP^3$ as an
example. Its fan consists of the cones
\begin{equation}
  \begin{split}
    \Sigma_{\CP^3} = 
    \Big\{ &
    \langle 0 \rangle,\;
    \langle e_1 \rangle,\;
    \langle e_2 \rangle,\;
    \langle e_3 \rangle,\;
    \langle -{\textstyle \sum} e_i \rangle,\;
    \\ &
    \langle e_1,e_2 \rangle,\;
    \langle e_1,e_3 \rangle,\;
    \langle e_1,-{\textstyle \sum} e_i \rangle,\;
    \langle e_2,e_3 \rangle,\;
    \langle e_2,-{\textstyle \sum} e_i \rangle,\;
    \langle e_3,-{\textstyle \sum} e_i \rangle,\;
    \\ &
    \langle e_1,e_2,e_3 \rangle,\;
    \langle e_1,e_2,-{\textstyle \sum} e_i \rangle,\;
    \langle e_1,e_3,-{\textstyle \sum} e_i \rangle,\;
    \langle e_2,e_3,-{\textstyle \sum} e_i \rangle,\;
    \Big\}
    ,
  \end{split}
\end{equation}
where $e_1$, $e_2$, $e_3$ are a basis for $N\simeq \Z^3$. There are
$4$ one-dimensional rays satisfying a unique linear relation, which
translates into $4$ homogeneous coordinates with the usual
identification
\begin{equation}
  \big[x_0:x_1:x_2:x_3\big] = 
  \big[\lambda x_0:\lambda x_1:\lambda x_2:\lambda x_3\big] 
  ,\quad
  \lambda \in \C^\times
  .
\end{equation}
A map of fans is a map of ambient lattices such that every cone of the
domain maps into a cone of the range fan. Any such fan morphism
defines a morphism of toric varieties in a covariantly functorial
way. For the purposes of this paper,\footnote{Except for
  \autoref{sec:basefib}.} we will only consider the case where the
lattice map is the identity. In this case the domain fan is simply a
subdivision of the range fan. The toric map corresponding to a
subdivision of a cone $\sigma$ is the blow-up along a toric subvariety
of dimension equal to $\codim(\sigma)$.

A toric divisor \footnote{All divisors in this paper will be Cartier
  divisors, even though we will be working with auxiliary singular
  varieties where not all divisors are Cartier.} is a formal linear
combination $D=\sum a_i V(x_i)$ of the codimension-one subvarieties
\begin{equation}
  V(x_i) = \big\{ x_i=0 \big\} 
\end{equation}
corresponding to the one-dimensional cones of the fan. There are two
basic constructions associated to such a toric divisor that will be
important in the following:
\begin{itemize}
\item Every coefficient $a_i$ can be thought of as the value of a
  function $f:N\to \Z$ on the generating lattice point $\rho_i$ of the
  $i$-th one-cone. If every cone is simplicial, then there is a
  uniquely defined continuous function on the fan with the above
  property. The pull-back of the function on the fan corresponds to
  the pull-back of the divisor by the toric map.
\item The divisor also defines a polytope
  \begin{equation}
    P_D = 
    \big\{ m \in M_\R \;\big|\; 
    \langle m, \rho_i \rangle \geq -a_i 
    \big\}
    ,
  \end{equation}
  where $M=N^\dual$ is the dual lattice. The global sections $\Gamma
  \Osheaf(D)$ are in one-to-one correspondence with the integral
  lattice points $M \cap P_D$ and can easily be counted for any given
  divisor.
\end{itemize}
A particularly relevant divisor is the anti-canonical divisor $-K=\sum
V(x_i)$. Given a polytope $\nabla \in N$, we can construct its face
fan and the polytope $\Delta = P_{-K}\subset M$. If $\Delta$ is again
a lattice polytope, then $\nabla$ is called reflexive.

Finally, note that $H^2(\CP^3)=\Z$. Hence, the line bundles on $\CP^3$
are classified by a single integer, their first Chern class. The toric
divisors, on the other hand, are defined by $4$ integers. Clearly,
there is no one-to-one correspondence between divisors $D$ and the
isomorphism class of the associated line bundle $\Osheaf(D)$. To make
this into a bijection, one must mod out \emph{linear equivalence} of
divisors. That is, one has to identify the piecewise linear functions
modulo linear functions. In particular, one can easily see that
\begin{equation}
  \label{eq:linequivP3}
  D=
  \sum_{i=0}^3
  a_i V(x_i)
  ~\sim~
  (a_0+a_1+a_2+a_3) V(x_0)
\end{equation}
on $\CP^3$.

\subsection{Three Birational Models}
\label{sec:three}

We now begin with the core of this paper and define the base threefold
of the elliptically fibered Calabi-Yau fourfold. In fact, I will
choose a smooth toric variety $\Bsmooth$ as the base manifold,
containing a non-simply connected divisor $\Dsmooth$. However,
directly analyzing $\Bsmooth$ will be overly complicated. In
particular, $\Bsmooth$ contains exceptional divisors that do not
intersect the divisor $\Dsmooth$ we are interested in. Therefore, to
better understand $\Dsmooth \subset \Bsmooth$, I will blow-down these
additional exceptional divisors. This will produce a singular variety
$\Bmed$ containing the same divisor $\Dmed = \Dsmooth$. Finally, I
will blow-down two more curves in $\Bmed$ to obtain an (even more
singular) three-dimensional variety $\Bsing$. The blown-down divisor
$\Dsing \subset \Bsing$ is the most suitable one to compute the
fundamental groups. To summarize, I am going to define successive
blow-ups
\begin{equation}
  \Bsmooth
  \stackrel{\pismooth}{\longrightarrow}
  \Bmed
  \stackrel{\pising}{\longrightarrow}
  \Bsing
\end{equation}
of three-dimensional toric varieties. Both of the maps $\pismooth$,
$\pising$ are toric morphisms defined in the obvious way by combining
cones of the fan into bigger cones, discarding all rays that are no
longer part of the more coarse (blown-down) fan.

\begin{figure}
  \centering
    \includegraphics{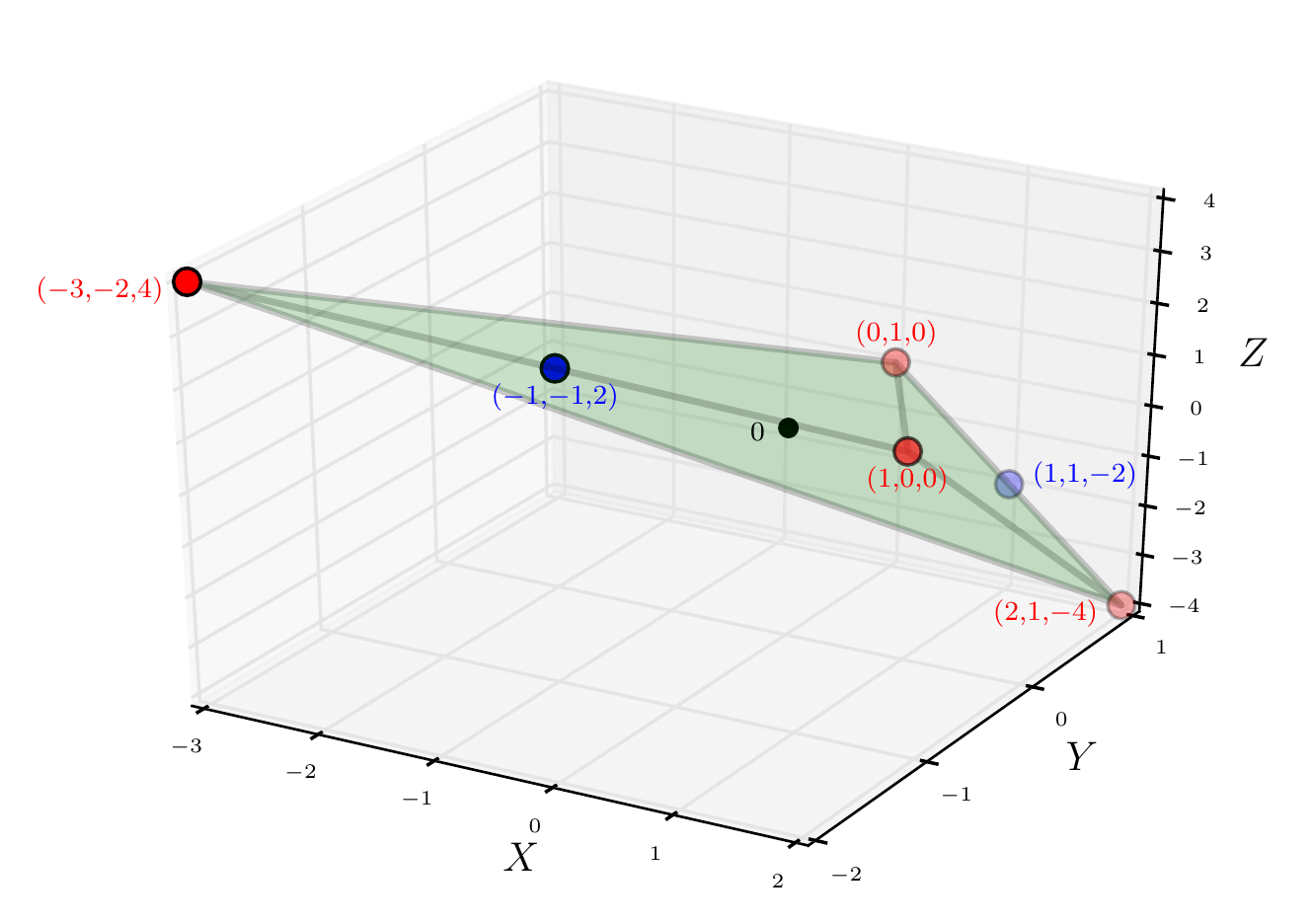}
    \caption{The rays $\Sigmaone_\Bsing$ (red dots) and 
      $\Sigmaone_\Bmed$ (red and blue dots).}
  \label{fig:NablaBsing}
\end{figure}
I now define the fans $\Sigma$ corresponding to the toric
varieties. Let me start by the rays $\Sigmaone$. The most singular
variety has
\begin{equation}
  \Sigmaone_\Bsing = 
  \big\{
  (-3, -2, 4),\  
  (0, 1, 0),\ 
  (1, 0, 0),\  
  (2, 1, -4)
  \big\},
\end{equation}
see \autoref{fig:NablaBsing}. The convex hull of these four points is
a tetrahedron, but not a minimal lattice simplex. In addition to the
origin (which is an interior point), it contains the two points $(-1,
-1, 2)$ and $(1, 1, -2)$ along two different edges. The variety
$\Bmed$ will be the maximal crepant partial resolution of $\Bsing$,
that is, the (in this case unique) maximal triangulation of the convex
hull $\conv(\Sigmaone_\Bsing)$. Hence, one must add the additional
integral points to the ray generators,
\begin{equation}
  \Sigmaone_\Bmed = 
  \Sigmaone_\Bsing \cup 
  \big\{
  (-1, -1, 2),\ 
  (1, 1, -2)
  \big\},
\end{equation}
Neither the variety $\Bsing$ nor its maximal crepant partial
resolution $\Bmed$ are smooth, related to the fact that the polytope
$\conv(\Sigmaone_\Bsing) = \conv(\Sigmaone_\Bmed)$ is not
reflexive. One again needs to add rays to resolve all singularities,
however this time the generators are necessarily outside of
$\conv(\Sigmaone_\Bsing)$. One particular choice I am going to make
are the rays generated by the $18$ points listed in
\autoref{tab:rays}.
\begin{table}
  \centering
  \renewcommand{\arraystretch}{1.3}
  \begin{tabular}{c|rrrrrrrrrrrrrrrrrr}
    $i$ & 
    0 & 1 & 2 & 3 & 4 & 5 & 6 & 7 & 8 & 9 & 
    10 & 11 & 12 & 13 & 14 & 15 & 16 & 17
    \\\hline
    \multirow{3}{*}{
      \begin{sideways}
        $i$-th ray
      \end{sideways}
    }
    & -3 & 0 & 1 & 2 & -1 & 1
    & -2 & -2 & -1 & -1 & -1 & 0 & 0 & 1 & 1 & 1 & 2 & 2 \\
    & -2 & 1 & 0 & 1 & -1 & 1
    & -1 & -1 & -1 & 0 & 0 & 0 & 0 & 0 & 0 & 1 & 1 & 1 \\
    & 4 & 0 & 0 & -4 & 2 & -2
    & 2 & 3 & 1 & 1 & 2 & -1 & 1 & -2 & -1 & -1 & -3 & -2
    \\ \hline
    $\Bsing$ &
    $\zsing_0$ & $\zsing_1$ & $\zsing_2$ & $\zsing_3$ 
    \\
    $\Bmed$ &
    $\zmed_0$ & $\zmed_1$ & $\zmed_2$ & $\zmed_3$ &
    $\zmed_4$ & $\zmed_5$ 
    \\
    $\Bsmooth$ &
    $\zsmooth_0$ & $\zsmooth_1$ & $\zsmooth_2$ & $\zsmooth_3$ & $\zsmooth_4$ & 
    $\zsmooth_5$ & $\zsmooth_6$ & $\zsmooth_7$ & $\zsmooth_8$ & $\zsmooth_9$ & 
    $\zsmooth_{10}$ & $\zsmooth_{11}$ & $\zsmooth_{12}$ &
    $\zsmooth_{13}$ & $\zsmooth_{14}$ & $\zsmooth_{15}$ & 
    $\zsmooth_{16}$ & $\zsmooth_{17}$ 
  \end{tabular}
  \caption{Ray generators and the associated Cox homogeneous variables.}
  \label{tab:rays}
\end{table}
\begin{figure}
  \centering
    \includegraphics{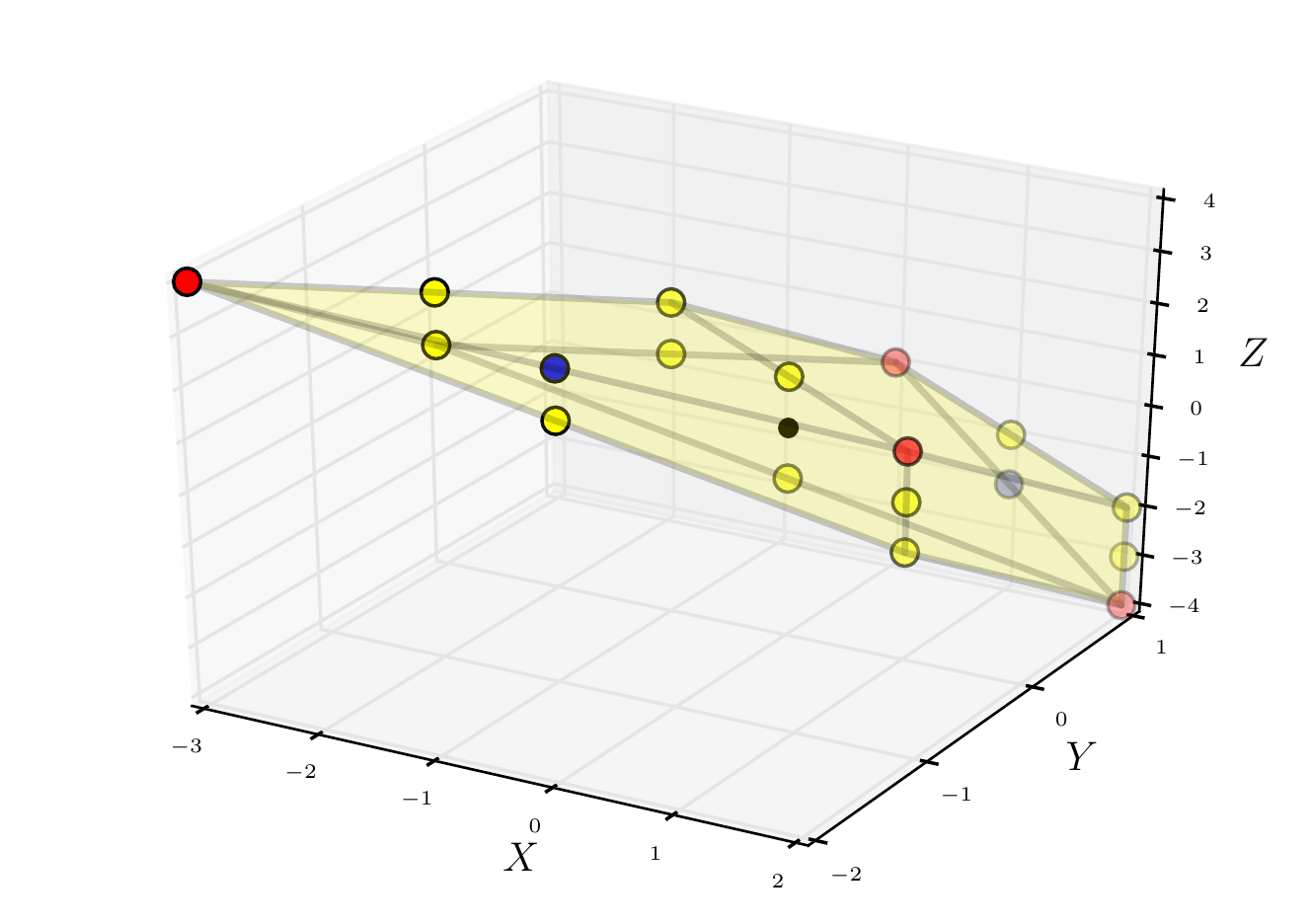}
  \caption{The rays $\Sigmaone_\Bsmooth$.}
  \label{fig:NablaBsmooth}
\end{figure}

The $8$ facets of the convex hull $\conv(\Sigmaone_\Bsmooth)$ are
given by the inequalities
\begin{equation}
  (n_x,\, n_y,\ n_z) \cdot
  \left(\begin{array}{rrrrrrrr}
      -1 & -1 & -1 & -1 & 0 & 1 & 1 & 2 \\
      -1 & 0 & 1 & 2 & -1 & -1 & 1 & -1 \\
      -1 & -1 & 0 & 0 & 0 & 0 & 1 & 1
    \end{array}\right)
  \geq
  (1, 1, 1, 1, 1, 1, 1, 1)
  ,
\end{equation}
which is, therefore, a reflexive polytope.

\begin{figure}
  \centering
    \includegraphics{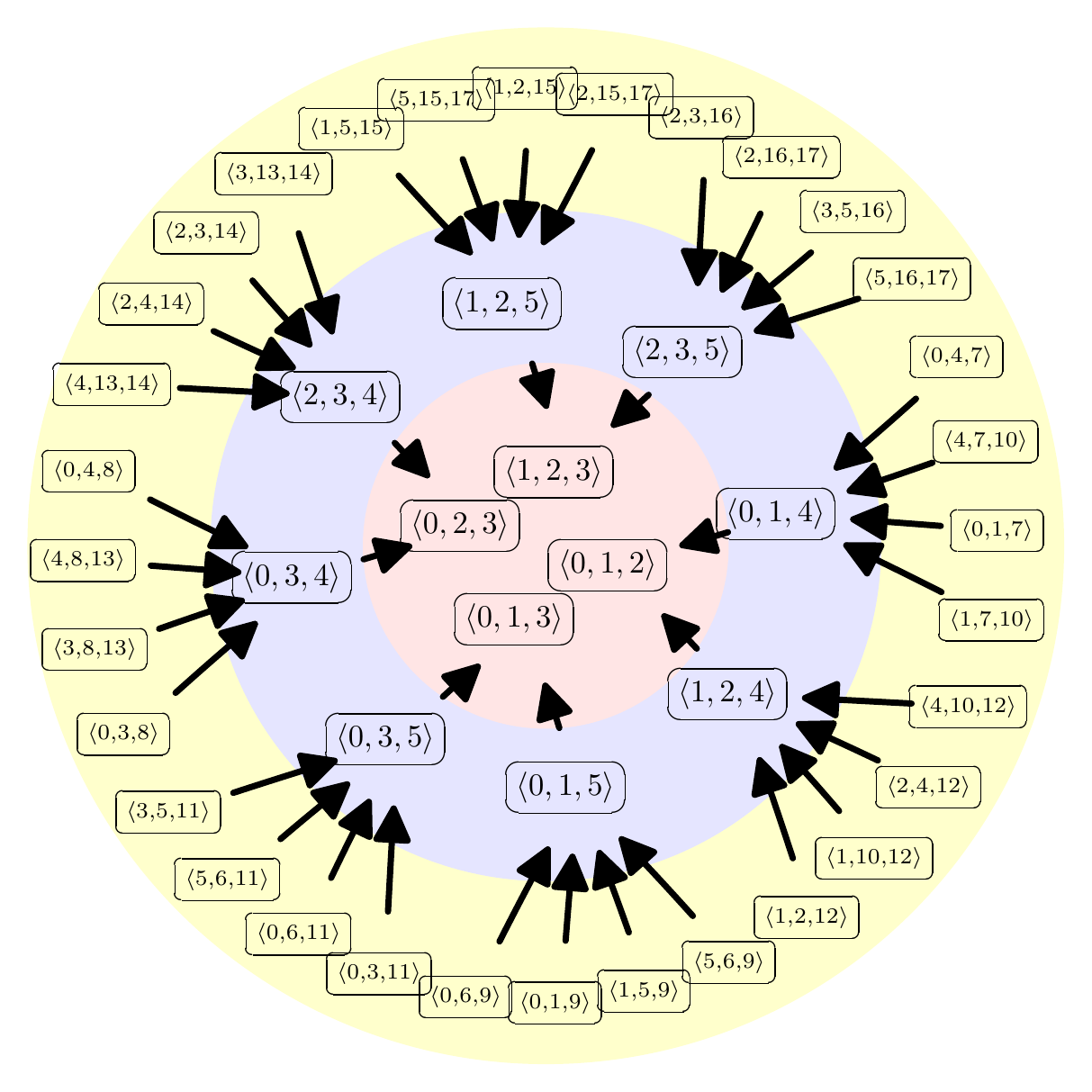}
    \caption[The generating cones of the fans $\Sigma_\Bsmooth$,
    $\Sigma_\Bmed$, and $\Sigma_\Bsing$.]{The generating cones of the
      fan $\Sigma_\Bsmooth$ (yellow, outer circle), $\Sigma_\Bmed$
      (blue), and $\Sigma_\Bsing$ (red, inner circle). By $\langle
      i,j,k \rangle$ we denote the cone spanned by the rays number
      $i$, $j$, and $k$ in \autoref{tab:rays}, and arrows mean
      ``contained in''.}
  \label{fig:fans_graph}
\end{figure}
Finally, to completely specify the toric varieties $\Bsing$, $\Bmed$,
and $\Bsmooth$, let me define the generating cones of the respective
fans:
\begin{itemize}
\item $\Sigma_\Bsing$ is the face fan of the polytope
  $\conv(\Sigmaone_\Bsing)$, see \autoref{fig:NablaBsing}.
\item $\Sigma_\Bmed$ is the unique maximal subdivision of
  $\Sigma_\Bsing$.
\item $\Sigma_\Bsmooth$ is a maximal subdivision of the face fan of
  the polytope $\conv(\Sigmaone_\Bsmooth)$, see
  \autoref{fig:NablaBsmooth}.
\end{itemize}
As there are many different maximal subdivisions of the face fan, this
alone does not uniquely specify the fan $\Sigma_\Bsmooth$. For
concreteness, I will fix the one listed in
\autoref{fig:fans_graph}. Note that not all combinatorial symmetries
of the graph in \autoref{fig:fans_graph} are actually symmetries of
the fan.

\subsection{Homogeneous Coordinates}
\label{sec:cox}

For future reference, let me list the toric Chow
groups~\cite{MR1234037}:
\begin{equation}
  \label{eq:Chow}
  A_k(\Bsmooth) = 
  \begin{cases}
    1 \\
    \Z^{15} \\
    \Z^{15} \\
    1 \\
  \end{cases}
  ,\quad
  A_k(\Bmed) = 
  \begin{cases}
    1 \\
    \Z^3 \times \Z_2 \\
    \Z^3 \times \Z_2^4 \\
    1 \\
  \end{cases}
  ,\quad
  A_k(\Bsing) = 
  \begin{cases}
    1                  & k=3 \\
    \Z \times \Z_4     & k=2  \\
    \Z \times \Z_2^2    & k=1 \\
    1                  & k=0 \\
  \end{cases}
\end{equation}
Since all three toric varieties have at most orbifold singularities,
the Hodge numbers are $h^{p,p} = \rank A_p$ and $h^{p,q}=0$ if $p\not=
q$.

The appearance of torsion in the Chow group slightly complicates the
Cox homogeneous coordinate~\cite{MR1299003} construction of the toric
varieties, so let me spell out the details. In general, a
simplicial\footnote{That is, with at most orbifold singularities.}
$d$-dimensional toric variety $X$ can be written as a geometric
quotient
\begin{equation}
  \label{eq:TVquot}
  X = 
  \frac{
    \C^{\Sigmaone_X} - Z
  }{
    \Hom\big(A_{d-1}(X),\; \C^\times\big)
  }
  \simeq
  \frac{
    \C^r - Z
  }{
    (\C^*)^{n-r}
    \times
    A_{d-1}(X)_\tors
  }
  ,
\end{equation}
where $r$ is the number of rays in the fan $\Sigma_X$. The exceptional
set $Z$ is the variety defined by the irrelevant ideal. A more catchy
way of remembering $Z$ is that it forbids homogeneous coordinates from
vanishing simultaneously if and only if their product is a monomial in
the Stanley-Reisner ideal. The latter is
\begin{equation}
  \label{eq:SR}
  \begin{split}
    SR(\Sigma_\Bsing) =&\;
    \big<
    \zsing_0 \zsing_1 \zsing_2 \zsing_3
    \big>
    ,
    \\
    SR(\Sigma_\Bmed) =&\;
    \big<
    \zmed_0 \zmed_2,\ 
    \zmed_1 \zmed_3,\ 
    \zmed_4 \zmed_5
    \big>
    ,
    \\
    SR(\Sigma_\Bsmooth) =&\;
    \big<
    \cdots
    \text{105 quadric monomials}
    \cdots
    \big>
    .
  \end{split}
\end{equation}
It remains to describe the groups in the denominator of
eq.~\eqref{eq:TVquot}. For the most singular variety $\Bsing$, one finds
\begin{equation}
  \begin{split}
    \label{eq:BsingG}
    \big[\zsing_0:\zsing_1:\zsing_2:\zsing_3\big]
    =&\;
    \big[
    \lambda \zsing_0:\lambda \zsing_1:
    \lambda \zsing_2:\lambda \zsing_3
    \big]
    \quad \forall \lambda \in \C^\times
    ,\\
    \big[\zsing_0:\zsing_1:\zsing_2:\zsing_3\big]
    =&\;
    \big[
    \zsing_0:
    \mu \zsing_1:
    \mu^2 \zsing_2:
    \mu^3 \zsing_3
    \big]
    \quad \forall 
    \mu \in \{1,i,i^2,i^3\} \simeq \Z_4
  \end{split}
\end{equation}
and for the intermediate blow-up $\Bmed$ 
\begin{equation}
  \begin{split}
    \label{eq:BmedG}
    \big[\zmed_0:\zmed_1:\zmed_2:\zmed_3:\zmed_4:\zmed_5\big]
    =&\;
    \big[
    \lambda \zmed_0:\lambda \zmed_1:
    \lambda \zmed_2:\lambda \zmed_3:
    \zmed_4:\zmed_5
    \big]
    \quad \forall \lambda \in \C^\times
    ,\\
    \big[\zmed_0:\zmed_1:\zmed_2:\zmed_3:\zmed_4:\zmed_5\big]
    =&\;
    \big[
    \lambda \zmed_0: \zmed_1:
    \lambda \zmed_2: \zmed_3:
    \lambda^{-2}\zmed_4:\zmed_5
    \big]
    \quad \forall \lambda \in \C^\times
    ,\\
    \big[\zmed_0:\zmed_1:\zmed_2:\zmed_3:\zmed_4:\zmed_5\big]
    =&\;
    \big[
    \zmed_0:\zmed_1:
    \zmed_2:\zmed_3:
    \lambda \zmed_4:\lambda \zmed_5
    \big]
    \quad \forall \lambda \in \C^\times
    ,\\
    \big[\zmed_0:\zmed_1:\zmed_2:\zmed_3:\zmed_4:\zmed_5\big]
    =&\;
    \big[
    \mu\zmed_0:
    \mu\zmed_1:
    \zmed_2:
    \zmed_3:
    \mu\zmed_4:
    \zmed_5
    \big]
    \quad \forall 
    \mu \in \{1,-1\} \simeq \Z_2
    .
  \end{split}
\end{equation}

\subsection{A Non-Simply Connnected Divisor}
\label{sec:divisor}

I am now going to define a divisor $\Dsing \subset \Bsing$ in the same
linear system as the toric divisor\footnote{$V(\zsing_i)$ denotes the
  toric divisor $\{\zsing_i=0\}$ associated to the $i$-th
  ray.}\footnote{By $\sim$, we will always denote rational equivalence
  of divisors. That is, $D_1\sim D_2$ means that there is a
  one-parameter family of divisors interpolating between $D_1$ and
  $D_2$. Equivalently, the Chow cycle defined by $D_1$ and $D_2$ is
  the same.}
\begin{equation}
  \Dsing
  \sim 4 V(\zsing_0)
  ,
\end{equation}
that is, as the zero set of a sufficiently generic section of the line
bundle $\Osheaf(\Dsing) = \Osheaf\big(V(\zsing_0)\big)^4$. A basis for
the sections is
\begin{equation}
  \label{eq:DsingSections}
  H^0\big(\Bsing, \Osheaf(\Dsing)\big) 
  =
  \Big<
  \zsing_0^4,\  
  \zsing_2^4,\  
  \zsing_1^4,\  
  \zsing_3^4,\  
  \zsing_1 \zsing_2^2 \zsing_3,\  
  \zsing_1^2 \zsing_3^2,\  
  \zsing_0 \zsing_2 \zsing_3^2,\  
  \zsing_0 \zsing_1^2 \zsing_2,\  
  \zsing_0^2 \zsing_2^2,\  
  \zsing_0^2 \zsing_1 \zsing_3
  \Big>,
\end{equation}
corresponding to the points of the polytope
\begin{equation}
  P_\Dsing = 
  \conv\big\{
  (0, 0, 0),\ 
  (4, 0, 2),\ 
  (0, 4, 1),\ 
  (0, 0, -1)
  \big\}
  \subset M
  .
\end{equation}
Note that the fan $\Sigma_\Bsing$ is precisely the normal fan of the
Newton polytope $P_\Dsing$. In this sense, $\Bsing$ is the ``natural''
ambient toric variety for the surface $\Dsing$.

For explicitness, let me fix once and for all a linear combination of
the monomials as the defining equation of the divisor $\Dsing$. I will
select the vertices of the Newton polytope and define
\begin{equation}
  \label{eq:Dsing}
  \Dsing = 
  \Big\{
  \zsing_0^4 + 
  \zsing_1^4 + 
  \zsing_2^4 + 
  \zsing_3^4 
  = 
  0
  \Big\}
  \subset 
  \Bsing
  .
\end{equation}
This surface is known to be an Enriques surface since it projects out
the potential $(2,0)$-form as mentioned in \autoref{sec:fore}. In
fact, this example has been known for some time, see Remark 3.6 in
\cite{MR1225527}.

\subsection{K\"ahler Cone and Canonical Divisors}
\label{eq:Kc}

The content of this subsection is not necessary for the understanding
of the paper, but I would like to pause for a moment and mention how
the ``Fermat quartic'' in eq.~\eqref{eq:Dsing} fails to define a $K3$
surface. In other words, how does the divisor $\Dsing = 4V(\zsing_0)$
differ from the anticanonical divisor
\begin{equation}
  -K_\Bsing = 
  V(\zsing_0) + V(\zsing_1) + V(\zsing_2) + V(\zsing_3)
\end{equation}
of $\Bsing$? Comparing with $\CP^3$, see eq.~\eqref{eq:linequivP3},
one might have thought that they were linearly equivalent.

Similarly to the quartic $K3\subset \CP^3$, one can also define a
Calabi-Yau variety in $\Bsing$ as the zero locus of a section of the
anticanonical bundle.\footnote{After resolution of singularities, the
  anticanonical divisor will be a smooth 2-dimensional Calabi-Yau
  manifold, that is, again a $K3$ surface.} The available sections are
\begin{equation}
  \label{eq:KBsingSections}
  H^0\big(\Bsing, \Osheaf(-K_\Bsing)\big) 
  =
  \Big<
  \zsing_0^2 \zsing_3^2,\ 
  \zsing_0^3 \zsing_2,\ 
  \zsing_2^2 \zsing_3^2,\  
  \zsing_0 \zsing_2^3,\  
  \zsing_1 \zsing_3^3,\  
  \zsing_0 \zsing_1 \zsing_2 \zsing_3,\  
  \zsing_0^2 \zsing_1^2,\  
  \zsing_1^2 \zsing_2^2,\  
  \zsing_1^3 \zsing_3
  \Big>
  .
\end{equation}
Note that this differs from the sections of $\Dsing$, see
eq.~\eqref{eq:KBsingSections}. Therefore, the two divisor are
\emph{not} linearly equivalent. Nevertheless, $-K_\Bsing$ and $\Dsing$
are very close to being linearly equivalent. In fact, its easy to see
that they are in the same rational divisor class\footnote{The toric
  divisor class group of a variety $X$ is often written as
  $\mathop{Cl}(X)$.  I will not use this notation in the following,
  but opt for $A_{\dim(X)-1}(X)$ instead.} since the rational divisor
class group is one-dimensional, $\dim \big(A_2(\Bsing)\otimes_\Z
\R\big) = 1$. However, their difference is a $2$-torsion element in
the (integral) divisor class
\begin{equation}
  \label{eq:KDsing2tors}
  K_\Bsing + \Dsing \not= 0 
  ,\quad
  2(K_\Bsing + \Dsing) = 0 
  \quad
  \in A_2(\Bsing)
  \simeq 
  \Z \oplus \Z_4
  .
\end{equation}
The same is true on the crepant partial resolution, where
$K_\Bmed+\Dmed$ is again a 2-torsion element in $A_2(\Bmed)$. 

On the final smooth resolution $\Bsmooth$ the divisor class group
$A_2(\Bsmooth)=\Z^{15}$ is torsion free. However, the last blow-up
$\pismooth:\Bsmooth\to\Bmed$ is not crepant, so
\begin{equation}
  \pismooth^*(K_\Bmed) 
  \not=
  K_\Bsmooth
  .
\end{equation}
Therefore, the divisors $-K_\Bsmooth$ and
$\Dsmooth=\pismooth^*(\Dmed)$ are no longer in the same rational
equivalence class.

Finally, let me describe the K\"ahler cones of these varieties. First,
let me remind the reader that the K\"ahler cone of a toric variety is
an open rational polyhedral cone in the rational divisor class group
corresponding to the cone of convex piecewise linear support functions
on the fan. For the two singular varieties, one obtains
\begin{equation}
  \begin{aligned}
    \Kc(\Bsing) 
    =&
    \big< V(\zsing_0) \big>
    &
    \subset&~
    A_2(\Bsing)\otimes_\Z \R \simeq \R
    \\
    \Kc(\Bmed) 
    =&
    \big< 
    V(\zmed_0),\  
    V(\zmed_1),\  
    2V(\zmed_0)+V(\zmed_4)
    \big>
    &
    \subset&~
    A_2(\Bmed)\otimes_\Z \R \simeq \R^3
  \end{aligned}
\end{equation}
As the anticanonical class $-K_\Bsing$ is rationally equivalent to
$4V(\zsing_0)$, we see that 
\begin{itemize}
\item $\Bsing$ is a (singular) Fano variety.
\item $\Bmed$ is not Fano, but the anticanonical class is on the
  boundary of the K\"ahler cone. In other words, $-K_\Bmed$ is nef but
  not ample.
\end{itemize}
On the smooth blow-up $\Bsmooth$, the K\"ahler cone 
\begin{equation}
  \Kc(\Bsmooth) 
  \quad\subset~
  A_2(\Bsmooth)\otimes_\Z \R \simeq \R^{15}
\end{equation}
is rather complicated and we will refrain from listing it
explicitly. It is spanned by the origin and $169$ rays and has $20$
facets.\footnote{That is, 14-dimensional faces.} The anticanonical
divisor $-K_\Bsmooth$ as well as $\Dsmooth$ sit on the boundary of the
K\"ahler cone, that is, are nef but not ample. However, each satisfies
a different subset of $16$ out of the $20$ facet equations, so they
lie on different faces of the K\"ahler cone.

\subsection{Pull-Back Divisors}
\label{sec:pullback}

By the usual dictionary of toric geometry, the toric divisor $4
V(\zsing_0) \sim D$ corresponds to a continuous piecewise linear
function on $N_\R \simeq \R^3$. Explicitly, the function is
\begin{equation}
  \label{eq:plfn}
  f(n_x, n_y, n_z) =
  \begin{cases}
    n_z & \text{if } \vec{n}\in \langle 0, 1, 2 \rangle
    , \\
    -4 n_x -2 n_z & \text{if } \vec{n}\in \langle 0, 1, 3 \rangle
    , \\
    -4 n_y - n_z & \text{if } \vec{n}\in \langle 0, 2, 3 \rangle
    , \\
    0 & \text{if } \vec{n}\in \langle 1, 2, 3 \rangle 
    .
  \end{cases}
\end{equation}
The pull-back of this toric divisor by the toric morphisms $\pising$
and $\pising\circ\pismooth$ is simply given by the pull-back of the
piecewise linear function. Therefore,
\begin{equation}
  \begin{split}
    \Dsing \sim&~
    4 V(\zsing_0)
    ,
    \\
    \Dmed  \sim&~
    4 V(\zmed_0) + 2 V(\zmed_4)
    ,
    \\
    \Dsmooth \sim&~
    4 V(\zsmooth_0) + 2 V(\zsmooth_4) + 4 V(\zsmooth_6) + 
    3 V(\zsmooth_7) + 3 V(\zsmooth_8) + 2 V(\zsmooth_9) + 
    \\
    &~
    2 V(\zsmooth_{10}) + 2 V(\zsmooth_{11}) + V(\zsmooth_{12}) + 
    2 V(\zsmooth_{13}) + V(\zsmooth_{14})
    .
  \end{split}
\end{equation}
What is the exceptional set of the first blow-up $\pising$?  Recall
that it corresponds to the subdivisions along the 2-cones
\begin{equation}
  \langle 0,4 \rangle 
  \cup
  \langle 2,4 \rangle 
  \to 
  \langle 0,2 \rangle 
  ,\quad
  \langle 1,5 \rangle 
  \cup
  \langle 3,5 \rangle 
  \to 
  \langle 1,3 \rangle 
  ,
\end{equation}
see \autoref{fig:NablaBsing}. Therefore, $\pising$ is the blow-up
along two disjoint rational curves of $\Z_2$-singularities in
$\Bsing$. A standard intersection computation in the Chow
group\cite{MR1234037} yields that each curve intersects the divisor
$\Dsing$ in two points. Therefore, the proper transform of $\Dmed
\subset \Bmed$ is $\Dsing$ blown up in four points. The final blow-up
$\pismooth:\Bsmooth\to\Bmed$ does not further subdivide the
$2$-skeleton $\Sigma^{(2)}_\Bmed$ and, therefore, corresponds to the
blow-up of points in $\Bmed$. Any sufficiently generic divisor $\Dmed$
misses these blow-up points and, therefore, the surfaces $\Dmed$ and
$\Dsmooth$ are isomorphic. To summarize,
\begin{itemize}
\item $\Dsing$ is a singular Enriques surface with four
  $\Z_2$-orbifold points.
\item $\Dsmooth$ and $\Dmed$ are the same smooth Enriques surface
  after blowing up the orbifold points.
\item Since the blow-up at a point does change the fundamental group, we
  find that
  \begin{equation}
    \pi_1\big(\Dsing\big) =
    \pi_1\big(\Dmed\big) =
    \pi_1\big(\Dsmooth\big) =
    \Z_2
    .
  \end{equation}
\end{itemize}

It is important to remember that the actual divisor is a fixed
subvariety defined as the zero locus of an equation. To relate this
defining equation before and after the blow-up, it is instructional to
write the first blow-up map $\pising:\Bmed\to\Bsing$ explicitly in
terms of its action on homogeneous coordinates. One finds
\begin{equation}
  \label{eq:pising}
  \pising
  \big(\big[
  \zmed_0: \zmed_1: \zmed_2: \zmed_3: \zmed_4: \zmed_5 
  \big]\big)
  =
  \big[
  \zmed_0 \sqrt{\zmed_4} :
  \zmed_1 \sqrt{\zmed_5} : 
  \zmed_2 \sqrt{\zmed_4} : 
  \zmed_3 \sqrt{\zmed_5}
  \big]
\end{equation}
Note that this map is well-defined on the equivalence classes
eq.~\eqref{eq:BmedG} thanks to the identifications
eq.~\eqref{eq:BsingG}. One sees that, for example, the section
$\zsing_0^4$ corresponds to the section $\zmed_0^4\zmed_4^2$ under the
$\pising^*$ pull-back. I leave the analogous expression for
$\pismooth$ as an exercise to the reader.

To summarize, the equation for the divisor $\Dsing$ determines the
equation satisfied by the proper transforms $\Dmed$ and $\Dsmooth$ on
the blow-ups. They are
\begin{equation}
  \label{eq:DivisorEq}
  \begin{split}
    \Dsing =& \big\{
    \zsing_0^4 +
    \zsing_1^4 +
    \zsing_2^4 +
    \zsing_3^4 =0 \big\}
    ,\\
    \Dmed =& \big\{
    \zmed_0^4 \zmed_4^2 +
    \zmed_1^4 \zmed_5^2 +
    \zmed_2^4 \zmed_4^2 +
    \zmed_3^4 \zmed_5^2
    =0 \big\}
    ,\\
    \Dsmooth =& \big\{
    \zsmooth_0^4 \zsmooth_4^2 \zsmooth_6^4 \zsmooth_7^3 
    \zsmooth_8^3 \zsmooth_9^2 \zsmooth_{10}^2 \zsmooth_{11}^2 
    \zsmooth_{12} \zsmooth_{13}^2 \zsmooth_{14} +
    \zsmooth_1^4 \zsmooth_5^2 \zsmooth_6^2 \zsmooth_7^2 
    \zsmooth_9^3 \zsmooth_{10}^4 \zsmooth_{11} \zsmooth_{12}^2 
    \zsmooth_{15}^3 \zsmooth_{16} \zsmooth_{17}^2 +
    \\ &\phantom{\big\{}
    \zsmooth_2^4 \zsmooth_4^2 \zsmooth_7 \zsmooth_8 
    \zsmooth_{10}^2 \zsmooth_{12}^3 \zsmooth_{13}^2 \zsmooth_{14}^3 
    \zsmooth_{15}^2 \zsmooth_{16}^2 \zsmooth_{17}^4 +
    \zsmooth_3^4 \zsmooth_5^2 \zsmooth_6^2 \zsmooth_8^2 
    \zsmooth_9 \zsmooth_{11}^3 \zsmooth_{13}^4 \zsmooth_{14}^2 
    \zsmooth_{15} \zsmooth_{16}^3 \zsmooth_{17}^2 
    =0 \big\}.
  \end{split}
\end{equation}


\section{Elliptic Fibration}
\label{sec:fibration}

So far, I have constructed 
\begin{itemize}
\item a three-dimensional (singular) Fano variety $\Bsing$,
\item a quasi-smooth divisor $\Dsing$ in $\Bsing$ with
  $\pi_1(\Dsing)=\Z_2$,
\item a smooth three-dimensional toric variety $\Bsmooth$,
  corresponding to a maximal subdivision of a reflexive polytope, and
\item a smooth divisor $\Dsmooth$ in $\Bsmooth$ with
  $\pi_1(\Dsmooth)=\Z_2$. This divisor is a smooth Enriques surface.
\end{itemize}
I will now proceed and construct four-dimensional elliptically fibered
Calabi-Yau varieties $\Ysing$, $\Ysmooth$ over $\Bsing$ and $\Bsmooth$
whose discriminant contains $\Dsing$ and $\Dsmooth$, respectively.

\subsection{Weierstrass Models}
\label{sec:Weierstrass}

Ideally, one would like to classify all elliptic fibrations over the
base manifold. Unfortunately it is not known how to do so in this
generality. It is known, however, that there exists a Weierstrass
model (not necessarily over the same base) which is a (in general)
different elliptic fibration~\cite{MR1929795, MR1918053}, at least
assuming that the base is smooth and the discriminant is a normal
crossing divisor. The Weierstrass model and the original elliptic
fibration are birational to each other, but apart from that their
relationship is arduous at best.

Having said this, let us define the elliptically fibered variety $Y$
in the most unimaginative way possible as a (global) Calabi-Yau
Weierstrass model
\begin{equation}
  Y
  =
  \Big\{
  y^2 z = x^3 + f(\vec \zeta) x z^2 + g(\vec \zeta) z^3
  \Big\}
  ~\subset~
  \mathbb{P}\Big( 
  \Osheaf
  \oplus
  \Osheaf(-2 K_Z) 
  \oplus
  \Osheaf(-3 K_Z)
  \Big)
  ,
\end{equation}
on a base variety $Z$ with coordinates $\zeta$. The remaining
coordinates, $z$, $x$, and $y$ are sections
\begin{equation}
  z\in \Gamma
  \Osheaf
  ,\quad
  x\in\Gamma
  \Osheaf(-2 K_Z)
  ,\quad
  y\in\Gamma
  \Osheaf(-3 K_Z)
  .
\end{equation}
The defining data of the Weierstrass model is the choice of
coefficients in the Weierstrass equation, that is, the choice of
sections
\begin{equation}
  f \in\Gamma 
  \Osheaf(-4 K_Z)
  ,\quad
  g \in\Gamma 
  \Osheaf(-6 K_Z)
\end{equation}
To engineer gauge theories on $7$-branes wrapped on a divisor $\{
\zeta = 0 \} \subset Z$, one needs suitable singularities. In
addition, the singularity must be of the correct split or non-split
type as in Tate's algorithm~\cite{MR0393039}. For this purpose it is
convenient to parametrize the Weierstrass\footnote{Technically, the
  singularity appears after blowing down all fiber components of the
  Weierstrass model not intersecting the zero section, but we will not
  dwell on this.} model by polynomials (that is, sections of suitable
line bundles) $a_1$, $a_2$, $a_3$, $a_4$, $a_6$ as
\begin{equation}
  \begin{split}
    f =&\; 
    -\tfrac{1}{48} a_1^4 
    - \tfrac{1}{6} a_1^2 a_2 
    + \tfrac{1}{2} a_1 a_3
    - \tfrac{1}{3} a_2^2 
    + a_4 
    \\
    g =&\; 
    \tfrac{1}{864} a_1^6 
    + \tfrac{1}{72} a_1^4 a_2 
    - \tfrac{1}{24} a_1^3 a_3 
    + \tfrac{1}{18} a_1^2 a_2^2 
    - \tfrac{1}{12} a_1^2 a_4 
    \\ &\;
    - \tfrac{1}{6} a_1 a_2 a_3 
    + \tfrac{2}{27} a_2^3 
    - \tfrac{1}{3} a_2 a_4 
    +  \tfrac{1}{4}  a_3^2 
    +  a_6
    .
  \end{split}
\end{equation}
The degree of vanishing of the $a_\ell(\zeta)$ then
determines\footnote{Except for a few special cases that will be of no
  relevance for us.} the low-energy effective gauge theory,
see~\cite{Bershadsky:1996nh, Choi:2010nf}. For everything to be
globally defined, the $a_\ell$ need to be sections of
\begin{equation}
  a_\ell
  \;\in\;
  \Gamma \Osheaf\big( -\ell K_\Bsmooth \big)
  .
\end{equation}


\subsection{Weierstrass Model on the Singular Base}
\label{sec:WBsing}

\begin{table}
  \centering
  \begin{tabular}{r|rrrrrr}
    $\dim \Gamma\Osheaf(-\kappa K_\Dsing-\delta \Dsing)$
    & $\kappa=1$ & $\kappa=2$ & $\kappa=3$ & $\kappa=4$ & $\kappa=5$ & $\kappa=6$ \\
    \hline
    $\delta=0$& $ 9$ & $ 43$ & $115$ & $245$ & $445$ & $735$ \\
    $\delta=1$& $ 0$ & $ 10$ & $ 42$ & $116$ & $244$ & $446$ \\
    $\delta=2$& $ 0$ & $  1$ & $  9$ & $ 43$ & $115$ & $245$ \\
    $\delta=3$& $ 0$ & $  0$ & $  0$ & $ 10$ & $ 42$ & $116$ \\
    $\delta=4$& $ 0$ & $  0$ & $  0$ & $  1$ & $  9$ & $ 43$ \\
    $\delta=5$& $ 0$ & $  0$ & $  0$ & $  0$ & $  0$ & $ 10$ \\
    $\delta=6$& $ 0$ & $  0$ & $  0$ & $  0$ & $  0$ & $  1$ \\
    $\delta\geq 7$& $ 0$ & $  0$ & $  0$ & $  0$ & $  0$ & $  0$ \\
  \end{tabular}
  \caption{Number of sections of $\Osheaf(-\kappa K_\Dsing-\delta \Dsing)$.}
  \label{tab:DsingSections}
\end{table}
To engineer a $SU(5)$ gauge theory coming from a $7$-brane wrapped on
the divisor $\Dsing$, one needs a split $A_4$
singularity~\cite{Bershadsky:1996nh}. This translates into $a_\ell$
vanishing to degree $\ell-1$ on $\Dsing$. In other words, $a_\ell$
must be divisible by $\dsing^{\ell-1}=0$ where $\dsing$ is the
defining equation for the divisor $\Dsing$ as given in
eq.~\eqref{eq:DivisorEq}. Put yet differently,
\begin{equation}
  \frac{a_\ell}{\dsing^{\ell-1}}
  \;\in\;
  \Gamma \Osheaf\big( -k K_\Dsing - (\ell-1) \Dsing \big)
\end{equation}
The number of sections is tabulated in \autoref{tab:DsingSections};
Note how the rows repeat with periodicity $2$. This again follows from
the fact that $K_\Bsing$ and $\Dsing$ differ by 2-torsion in the
divisor class group, see eq.~\eqref{eq:KDsing2tors}. Hence, there are
plenty sections available for $a_1$, $\dots$, $a_6$ and one can easily
find an elliptic fibration with a split $A_4$ over $\Dsing$.

\subsection{Weierstrass Model on the Smooth Base}
\label{sec:WBsmooth}

Let me now turn to the smooth threefold $\Bsmooth$ and construct a
suitable singularity over the smooth divisor $\Dsmooth$.
\begin{table}
  \centering
  \begin{tabular}{r|rrrrrr}
    $\dim \Gamma\Osheaf(-\kappa K_\Dsmooth-\delta \Dsmooth)$
    & $\kappa=1$ & $\kappa=2$ & $\kappa=3$ & $\kappa=4$ & $\kappa=5$ & $\kappa=6$ \\
    \hline
    $\delta=0$& $  9$ & $ 35$ & $ 91$ & $189$ & $341$ & $559$ \\
    $\delta=1$& $  0$ & $  2$ & $ 18$ & $ 60$ & $140$ & $270$ \\
    $\delta=2$& $  0$ & $  0$ & $  0$ & $  3$ & $ 27$ & $ 85$ \\
    $\delta=3$& $  0$ & $  0$ & $  0$ & $  0$ & $  0$ & $  4$ \\
    $\delta\geq 4$& $  0$ & $  0$ & $  0$ & $  0$ & $  0$ & $  0$ \\
  \end{tabular}
  \caption[Number of sections of $\Osheaf(-\kappa K_\Dsmooth-\delta
  \Dsmooth)$]{Number of sections of $\Osheaf(-\kappa K_\Dsmooth-\delta
    \Dsmooth)$, see also \autoref{tab:DsingSections}.}
  \label{tab:DsmoothSections}
\end{table}
The main difference is that now, after resolving the singularity, the
anticanonical divisor is ``smaller'' than $\Dsmooth$, by which I mean
that there are strictly less sections available for the Weierstrass
model. See \autoref{tab:DsmoothSections} for details. Note that, if
one always imposes the maximal degree of vanishing such that there are
still non-zero sections, one can at most implement a split $A_2$
singularity leading to a low-energy $SU(3)$ gauge theory.

Having being dealt this lemon, let me try to make some lemonade. As in
the previous subsection, I will write $\Dsmooth=\{\dsmooth=0\}$ for
the defining equation, see eq.~\eqref{eq:DivisorEq}. The split $A_2$
singularity corresponds to a factorized form
\begin{equation}
  \begin{aligned}
    a_1 =&\; \alpha_1
    &\qquad
    \alpha_1 \in&\; \Gamma \Osheaf(-K_\Bsmooth)
    ,\\
    a_2 =&\; \dsmooth \alpha_2
    &
    \alpha_2 \in&\; \Gamma \Osheaf(-2K_\Bsmooth-\Dsmooth)
    ,\\
    a_3 =&\; \dsmooth \alpha_3
    &
    \alpha_3 \in&\; \Gamma \Osheaf(-3K_\Bsmooth-\Dsmooth)
    ,\\
    a_4 =&\; \dsmooth^2 \alpha_4
    &
    \alpha_4 \in&\; \Gamma \Osheaf(-4K_\Bsmooth-2\Dsmooth)
    ,\\
    a_6 =&\; \dsmooth^3 \alpha_6
    &
    \alpha_6 \in&\; \Gamma \Osheaf(-6K_\Bsmooth-3\Dsmooth)
    .
  \end{aligned}
\end{equation}
A basis for all sections can, of course, be written as in terms of
homogeneous monomials in the $18$ homogeneous coordinates
$\zsmooth_0$, $\dots$, $\zsmooth_{17}$. To save a tree I will now
switch to inhomogeneous coordinates $(\xi_0, \xi_1, \xi_2) \in \C^3$
for the coordinate patch, say, corresponding to the cone $\langle
0,4,7\rangle$. This amounts to replacing the homogeneous coordinates
with
\begin{equation}
  \big[\zsmooth_0:\zsmooth_1:\cdots:\zsmooth_{17}\big]
  = 
  \big[\xi_0:1:1:1:\xi_1:1:1:\xi_2:1:\cdots:1\big]
\end{equation}
In this patch,
\begin{equation}
  \dsmooth = 
  \xi_0^4 \xi_1^2 \xi_2^3 + \xi_1^2 \xi_2 + \xi_2^2 + 1
\end{equation}
and the sections of the relevant line bundles are 
\begin{equation}
  \begin{split}
    \Gamma \Osheaf(-K_\Bsmooth)
    =&\;
    \big\langle
    1, \xi_1, \xi_2, \xi_0 \xi_1^2 \xi_2, \xi_1 \xi_2, 
    \xi_0^2 \xi_1 \xi_2, \xi_0^2 \xi_1 \xi_2^2, 
    \xi_0^3 \xi_1^2 \xi_2^2, \xi_0 \xi_1 \xi_2
    \big\rangle
    ,\\
    \Gamma \Osheaf(-2K_\Bsmooth-\Dsmooth)
    =&\;
    \big\langle
    1, \xi_0^2 \xi_1^2 \xi_2
    \big\rangle
    ,\\
    \Gamma \Osheaf(-3K_\Bsmooth-\Dsmooth)
    =&\;
    \big\langle
    1, \xi_1, \xi_2, \xi_1 \xi_2, \xi_0^2 \xi_1^3 \xi_2, 
    \xi_0^3 \xi_1^4 \xi_2^2, \xi_0^2 \xi_1^3 \xi_2^2, 
    \xi_0^2 \xi_1 \xi_2, \xi_0^2 \xi_1 \xi_2^2, 
    \xi_0^4 \xi_1^3 \xi_2^2,
    \\&\;\phantom{\big\langle}
    \xi_0^4 \xi_1^3 \xi_2^3, 
    \xi_0^5 \xi_1^4 \xi_2^3, \xi_0 \xi_1^2 \xi_2, 
    \xi_0^2 \xi_1^2 \xi_2^2, \xi_0^3 \xi_1^3 \xi_2^2, 
    \xi_0 \xi_1 \xi_2, \xi_0^2 \xi_1^2 \xi_2, 
    \xi_0^3 \xi_1^2 \xi_2^2
    \big\rangle
    ,\\
    \Gamma \Osheaf(-4K_\Bsmooth-2\Dsmooth)
    =&\;
    \big\langle
    1, \xi_0^4 \xi_1^4 \xi_2^2, \xi_0^2 \xi_1^2 \xi_2
    \big\rangle
    ,\\
    \Gamma \Osheaf(-6K_\Bsmooth-3\Dsmooth)
    =&\;
    \big\langle
    1, \xi_0^6 \xi_1^6 \xi_2^3, \xi_0^2 \xi_1^2 \xi_2, \xi_0^4 \xi_1^4 \xi_2^2
    \big\rangle
    .
  \end{split}
\end{equation}
For simplicity I will choose $\alpha_\ell$ to be the sum of the
monomials corresponding to the vertices of the Newton polyhedron, that
is,
\begin{equation}
  \begin{split}
    \alpha_1 =&\;
    \xi_0^3 \xi_1^2 \xi_2^2 + \xi_0^2 \xi_1 \xi_2^2 + 
    \xi_0^2 \xi_1 \xi_2 + \xi_0 \xi_1^2 \xi_2 + 
    \xi_1 \xi_2 + \xi_1 + \xi_2 + 1
    ,\\
    \alpha_2 =&\;
    \xi_0^2 \xi_1^2 \xi_2 + 1
    ,\\
    \alpha_3 =&\;
    \xi_0^5 \xi_1^4 \xi_2^3 + \xi_0^4 \xi_1^3 \xi_2^3 + 
    \xi_0^4 \xi_1^3 \xi_2^2 + \xi_0^3 \xi_1^4 \xi_2^2 + 
    \\&\;
    \xi_0^2 \xi_1^3 \xi_2^2 + \xi_0^2 \xi_1^3 \xi_2 + 
    \xi_0^2 \xi_1 \xi_2^2 + \xi_0^2 \xi_1 \xi_2 + 
    \xi_1 \xi_2 + \xi_1 + \xi_2 + 1
    ,\\
    \alpha_4 =&\;
    \xi_0^4 \xi_1^4 \xi_2^2 + 1
    ,\\
    \alpha_6 =&\;
    \xi_0^6 \xi_1^6 \xi_2^3 + 1
    .
  \end{split}
\end{equation}
For all purposes in the following, this choice is generic. By
construction, the discriminant then factorizes as
\begin{equation}
  \Delta =
  4f^3+27g^2 =
  \dsmooth^3 \; \rsmooth
  ,
\end{equation}
where the remainder\footnote{In fact, $\rsmooth$ is a polynomial
  consisting of $1083$ monomials in $\xi_0$, $\xi_1$, and $\xi_2$.}
defines a new divisor $\Rsmooth \eqdef \{\rsmooth=0\}\sim
-12K_\Bsmooth-3\Dsmooth$. In particular, the homology class splits as
\begin{equation}
  [\Delta] = 3 [\Dsmooth] + [\Rsmooth]
\end{equation}
Using the explicit equations, one can check~\cite{GPS05} that
\begin{itemize}
\item $\Rsmooth$ is an irreducible divisor.
\item Neither $f$ nor $g$ vanish at a generic point of
  $\Rsmooth$. Hence it supports $I_0$ Kodaira fibers in the
  Weierstrass model.
\item $\Dsmooth$ is smooth.
\item $\Rsmooth$ is not smooth, for example
  $(\xi_0,\xi_1,\xi_2)=(1,1,-1)$ is a singular point.
\item The curve $\Dsmooth \cap \Rsmooth$ is not a complete
  intersection.
\end{itemize}

Let me further investigate the intersection curve $\Dsmooth \cap
\Rsmooth$. One component (in the $\langle 0,4,7\rangle$ patch) is
given by the surprisingly simple expression
\begin{equation}
  c: 
  \C \to \Dsmooth \cap \Rsmooth
  ,~
  t \mapsto (t,0,i)
  .
\end{equation}
Therefore, $(\xi_1,\xi_2)$ are good normal coordinates. Taylor
expanding along the normal directions for a generic point $c(t)$, we
see that $\Dsmooth$ and $\Rsmooth$ share the same tangent plane but do
not osculate to any higher degree. Therefore, the degree of vanishing
of the discriminant jumps from $5$ to $7$ along the intersection locus
$\Dsmooth\cap\Rsmooth$, corresponding to worsening of the $A_2$
singularity to an $A_4$ singularity.


\section{Conclusions}
\label{sec:conclusion}

In this paper I have constructed F-theory models with, a priori,
$SU(5)$ gauge theory on a singular Fano threefold and a $SU(3)$ gauge
theory on the blown-up smooth threefold. In both cases the
non-Abelian gauge theory comes from a $7$-brane wrapped on an Enriques
surface, which has fundamental group $\Z_2$. Therefore, in both cases
one can switch on a discrete Wilson line and break the gauge group
below the compactification scale in the usual manner.

The fact that the only partially resolved base $\Bmed$ allows for a
higher rank gauge group on the $7$-brane than its smooth blow-up is
curious: One might be tempted to interpret the K\"ahler deformation as
the usual Higgs mechanism, however the singular points are disjoint
from the $7$-brane. In any case, there must be further physical
degrees of freedom associated to the singularities in the base and it
would be nice to have a more concise F-theory dictionary for them.

\appendix

\section{Fibrations of the Base}
\label{sec:basefib}

\newcommand{\Shat}{{\hat{S}}}

\begin{figure}
  \centering
    \includegraphics{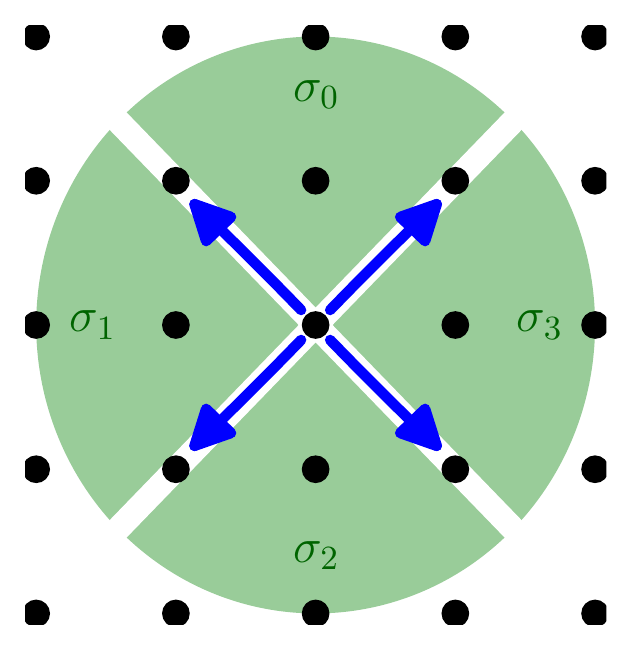}
    \hspace{1cm}
    \includegraphics{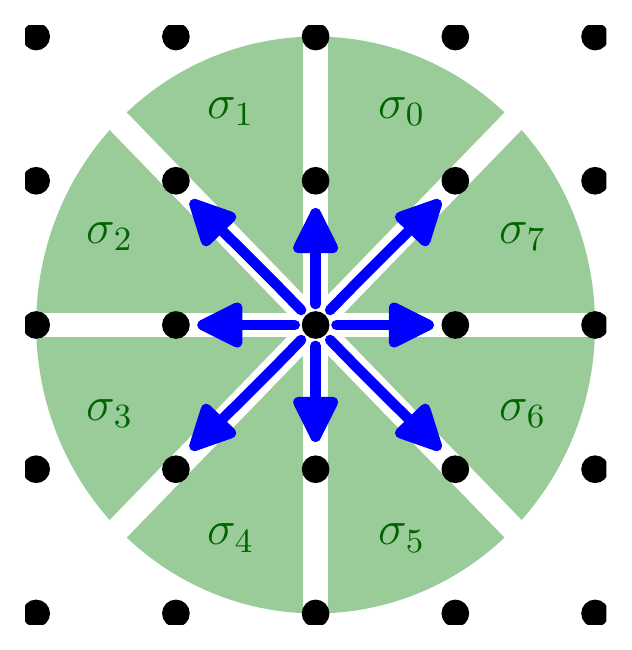}
    \caption[The fans of $\big(\CP^1\times \CP^1\big)/\Z_2$ and its
    blow-up.]{The fan defining the toric variety $S = \big(\CP^1\times
      \CP^1\big)/\Z_2$ (left) and its crepant smooth resolution
      $\Shat$ (right).}
  \label{fig:SShat}
\end{figure}
The base manifolds $\Bsing$, $\Bmed$, and $\Bsmooth$ are fibered in an
interesting manner which I will describe in this appendix. The map to
the $2$-dimensional base is given by the $N$-lattice projection
\begin{equation}
  \phi:~
  N^{(3)} \to N^{(2)}
  ,\quad
  \vec{n} \mapsto
  \begin{pmatrix}
     1& 1& 1\\
    -1& 1& 0
  \end{pmatrix}
  \vec{n}
\end{equation}
This defines a toric morphism of toric varieties if and only if every
cone of the domain fan is mapped into a cone of the range fan. It is
easy to see that the rays $\Sigmaone_\Bsing$ and $\Sigmaone_\Bmed$ map
to the rays of the fan of $S = \big(\CP^1\times \CP^1\big)/\Z_2$, and
the rays $\Sigmaone_\Bsmooth$ map to the crepant resolution
$\Shat$. See \autoref{fig:SShat} for a graphical representation of the
fans of $S$ and $\Shat$.

However, consistently mapping the rays of the fans is not enough to
define a toric morphism. Checking all higher-dimensional cones with
respect to the lattice homomorphism $\phi$, one finds that
\begin{itemize}
\item The variety $\Bsing$ is not fibered.
\item The variety $\Bmed$ is a $\CP^1$-fibrations over $S$.
\item The smooth threefold $\Bsmooth$ is a $\CP^1$-fibration over $S$,
  but not over the crepant resolution $\Shat$.
\end{itemize}
It is, perhaps, vexing that the resolved threefold $\Bsmooth$ is not a
fibration over the resolved base $\Shat$. However, a closer
investigation reveals that one can flop $4$ offending curves,
corresponding to the $4$ bistellar flips
\begin{equation}
  \begin{split}
    \big\{ \langle 1,2,15\rangle,\langle 1,2,12\rangle \big\}
    \mapsto&\;
    \big\{ \langle 1,12,15\rangle,\langle 2,12,15\rangle \big\}
    \\
    \big\{ \langle 0,1,9\rangle,\langle 0,1,7\rangle \big\}
    \mapsto&\;
    \big\{ \langle 0,7,9\rangle,\langle 1,7,9\rangle \big\}
    \\
    \big\{ \langle 0,3,11\rangle,\langle 0,3,8\rangle \big\}    
    \mapsto&\;
    \big\{ \langle 0,8,11\rangle,\langle 3,8,11\rangle \big\}
    \\
    \big\{ \langle 2,3,16\rangle,\langle 2,3,14\rangle \big\}  
    \mapsto&\;
    \big\{ \langle 2,14,16\rangle,\langle 3,14,16\rangle \big\}
  \end{split}
\end{equation}
of the fan $\Sigma_\Bsmooth$. The flopped threefold is then a $\CP^1$
fibration over the resolved base $\Shat$. Of course the flopped
threefold is then only birational to $\Bmed$, $\Bsing$ and no longer a
direct blow-up. However, it supports essentially the same elliptic
fibration as $\Bsmooth$ as constructed in \autoref{sec:fibration}.

\bibliographystyle{utcaps} 
\renewcommand{\refname}{Bibliography}
\addcontentsline{toc}{section}{Bibliography} 
\bibliography{Main}

\providecommand{\href}[2]{#2}\begingroup\raggedright\begin{thebibliography}{10}

\bibitem{Vafa:1996xn}
C.~Vafa, ``{Evidence for F-Theory},'' {\em Nucl. Phys.} {\bf B469} (1996)
  403--418,
\href{http://arXiv.org/abs/hep-th/9602022}{{\tt hep-th/9602022}}.

\bibitem{Sen:1996vd}
A.~Sen, ``{F-theory and Orientifolds},'' {\em Nucl. Phys.} {\bf B475} (1996)
  562--578,
\href{http://arXiv.org/abs/hep-th/9605150}{{\tt hep-th/9605150}}.

\bibitem{Beasley:2008dc}
C.~Beasley, J.~J. Heckman, and C.~Vafa, ``{GUTs and Exceptional Branes in
  F-theory - I},'' {\em JHEP} {\bf 01} (2009) 058,
\href{http://arXiv.org/abs/0802.3391}{{\tt 0802.3391}}.

\bibitem{Beasley:2008kw}
C.~Beasley, J.~J. Heckman, and C.~Vafa, ``{GUTs and Exceptional Branes in
  F-theory - II: Experimental Predictions},'' {\em JHEP} {\bf 01} (2009) 059,
\href{http://arXiv.org/abs/0806.0102}{{\tt 0806.0102}}.

\bibitem{Donagi:2008ca}
R.~Donagi and M.~Wijnholt, ``{Model Building with F-Theory},''
\href{http://arXiv.org/abs/0802.2969}{{\tt 0802.2969}}.

\bibitem{Marsano:2009ym}
J.~Marsano, N.~Saulina, and S.~Schafer-Nameki, ``{F-theory Compactifications
  for Supersymmetric GUTs},'' {\em JHEP} {\bf 08} (2009) 030,
\href{http://arXiv.org/abs/0904.3932}{{\tt 0904.3932}}.

\bibitem{Blumenhagen:2009yv}
R.~Blumenhagen, T.~W. Grimm, B.~Jurke, and T.~Weigand, ``{Global F-theory
  GUTs},'' {\em Nucl. Phys.} {\bf B829} (2010) 325--369,
\href{http://arXiv.org/abs/0908.1784}{{\tt 0908.1784}}.

\bibitem{Chen:2010ts}
C.-M. Chen, J.~Knapp, M.~Kreuzer, and C.~Mayrhofer, ``{Global SO(10) F-theory
  GUTs},''
\href{http://arXiv.org/abs/1005.5735}{{\tt 1005.5735}}.

\bibitem{Chen:2010tp}
C.-M. Chen and Y.-C. Chung, ``{Flipped SU(5) GUTs from E8 Singularity in
  F-theory},''
\href{http://arXiv.org/abs/1005.5728}{{\tt 1005.5728}}.

\bibitem{Blumenhagen:2010at}
R.~Blumenhagen, ``{Basics of F-theory from the Type IIB Perspective},'' {\em
  Fortsch. Phys.} {\bf 58} (2010) 820--826,
\href{http://arXiv.org/abs/1002.2836}{{\tt 1002.2836}}.

\bibitem{Hosotani1}
Y.~Hosotani, ``{Dynamical Mass Generation by Compact Extra Dimensions},'' {\em
  Phys. Lett.} {\bf B126} (1983)
309.

\bibitem{Hosotani2}
Y.~Hosotani, ``{Dynamics of Nonintegrable Phases and Gauge Symmetry
  Breaking},'' {\em Ann. Phys.} {\bf 190} (1989)
233.

\bibitem{Blumenhagen:2008zz}
R.~Blumenhagen, V.~Braun, T.~W. Grimm, and T.~Weigand, ``{GUTs in Type IIB
  Orientifold Compactifications},'' {\em Nucl. Phys.} {\bf B815} (2009) 1--94,
\href{http://arXiv.org/abs/0811.2936}{{\tt 0811.2936}}.

\bibitem{Donagi:2008kj}
R.~Donagi and M.~Wijnholt, ``{Breaking GUT Groups in F-Theory},''
\href{http://arXiv.org/abs/0808.2223}{{\tt 0808.2223}}.

\bibitem{Blumenhagen:2008aw}
R.~Blumenhagen, ``{Gauge Coupling Unification in F-Theory Grand Unified
  Theories},'' {\em Phys. Rev. Lett.} {\bf 102} (2009) 071601,
\href{http://arXiv.org/abs/0812.0248}{{\tt 0812.0248}}.

\bibitem{ToricVarieties}
V.~Braun and A.~Novoseltsev, ``Toric Geometry in the Sage CAS.'' to appear.

\bibitem{Sage}
W.~Stein {\em et al.}, {\em {S}age {M}athematics {S}oftware ({V}ersion 4.5.3)}.
\newblock The Sage Development Team, 2010.
\newblock {\tt http://www.sagemath.org}.

\bibitem{GPS05}
G.-M. Greuel, G.~Pfister, and H.~Sch\"onemann, ``{\sc Singular} 3.0,'' a
  computer algebra system for polynomial computations, Centre for Computer
  Algebra, University of Kaiserslautern, 2005.
\newblock \url{http://www.singular.uni-kl.de}.

\bibitem{Nahm:1999ps}
W.~Nahm and K.~Wendland, ``{A hiker's guide to K3: Aspects of N = (4,4)
  superconformal field theory with central charge c = 6},'' {\em Commun. Math.
  Phys.} {\bf 216} (2001) 85--138,
\href{http://arXiv.org/abs/hep-th/9912067}{{\tt hep-th/9912067}}.

\bibitem{MR2282962}
V.~Batyrev and M.~Kreuzer, ``Integral cohomology and mirror symmetry for
  {C}alabi-{Y}au 3-folds,'' in {\em Mirror symmetry. {V}}, vol.~38 of {\em
  AMS/IP Stud. Adv. Math.}, pp.~255--270.
\newblock Amer. Math. Soc., Providence, RI, 2006.

\bibitem{MR2122419}
B.~Nill, ``Gorenstein toric {F}ano varieties,'' {\em Manuscripta Math.} {\bf
  116} (2005), no.~2, 183--210.

\bibitem{Braun:2010vc}
V.~Braun, ``{On Free Quotients of Complete Intersection Calabi-Yau
  Manifolds},''
\href{http://arXiv.org/abs/1003.3235}{{\tt 1003.3235}}.

\bibitem{Braun:2009qy}
V.~Braun, P.~Candelas, and R.~Davies, ``{A Three-Generation Calabi-Yau Manifold
  with Small Hodge Numbers},'' {\em Fortsch. Phys.} {\bf 58} (2010) 467--502,
\href{http://arXiv.org/abs/0910.5464}{{\tt 0910.5464}}.

\bibitem{Candelas:2010ve}
P.~Candelas and A.~Constantin, ``{Completing the Web of $Z_3$ - Quotients of
  Complete Intersection Calabi-Yau Manifolds},''
\href{http://arXiv.org/abs/1010.1878}{{\tt 1010.1878}}.

\bibitem{MR1299003}
D.~A. Cox, ``The homogeneous coordinate ring of a toric variety,'' {\em J.
  Algebraic Geom.} {\bf 4} (1995), no.~1, 17--50.

\bibitem{MR1234037}
W.~Fulton, {\em Introduction to toric varieties}, vol.~131 of {\em Annals of
  Mathematics Studies}.
\newblock Princeton University Press, Princeton, NJ, 1993.
\newblock The William H. Roever Lectures in Geometry.

\bibitem{MR1225527}
M.~Oka, ``Finiteness of fundamental group of compact convex integral
  polyhedra,'' {\em Kodai Math. J.} {\bf 16} (1993), no.~2, 181--195.

\bibitem{MR1929795}
N.~Nakayama, ``Local structure of an elliptic fibration,'' in {\em Higher
  dimensional birational geometry ({K}yoto, 1997)}, vol.~35 of {\em Adv. Stud.
  Pure Math.}, pp.~185--295.
\newblock Math. Soc. Japan, Tokyo, 2002.

\bibitem{MR1918053}
N.~Nakayama, ``Global structure of an elliptic fibration,'' {\em Publ. Res.
  Inst. Math. Sci.} {\bf 38} (2002), no.~3, 451--649.

\bibitem{MR0393039}
J.~Tate, ``Algorithm for determining the type of a singular fiber in an
  elliptic pencil,'' in {\em Modular functions of one variable, {IV} ({P}roc.
  {I}nternat. {S}ummer {S}chool, {U}niv. {A}ntwerp, {A}ntwerp, 1972)},
  pp.~33--52. Lecture Notes in Math., Vol. 476.
\newblock Springer, Berlin, 1975.

\bibitem{Bershadsky:1996nh}
M.~Bershadsky {\em et al.}, ``{Geometric singularities and enhanced gauge
  symmetries},'' {\em Nucl. Phys.} {\bf B481} (1996) 215--252,
\href{http://arXiv.org/abs/hep-th/9605200}{{\tt hep-th/9605200}}.

\bibitem{Choi:2010nf}
K.-S. Choi, ``{SU(3) x SU(2) x U(1) Vacua in F-Theory},''
\href{http://arXiv.org/abs/1007.3843}{{\tt 1007.3843}}.

\end{thebibliography}\endgroup

\end{document}